\documentclass[a4paper]{article}
\usepackage{graphicx} % Required for inserting images
\usepackage{url}
\usepackage{comment}

\title{Software Reuse in the Generative AI Era: From Cargo Cult Towards AI Native Software Engineering}
\author{Tommi Mikkonen$^1$ and Antero Taivalsaari$^2$\\
$^1$University of Jyväskylä, Jyväskylä, Finland\\
$^2$Nokia Technologies, Tampere, Finland\\
tommi.j.mikkonen@jyu.fi, antero.taivalsaari@nokia.com
}
\date{June 2025}

\begin{document}

\maketitle

\noindent
\textbf{Abstract}
Software development is currently under a paradigm shift in which artificial intelligence and generative software reuse are taking the center stage in software creation. Consequently, earlier software reuse practices and methods are rapidly being replaced by AI-assisted approaches in which developers place their trust on code that has been generated by artificial intelligence.  This is leading to a new form of software reuse that is conceptually not all that different from cargo cult development. In this paper we discuss the implications of AI-assisted generative software reuse in the context of emerging "AI native" software engineering, bring forth relevant questions, and define a tentative research agenda and call to action for tackling some of the central issues associated with this approach.
\smallskip


\section{Introduction}

Software development is currently under a paradigm shift in which Artificial Intelligence (AI) -- in particular Generative AI \cite{ebert2023generative} -- has taken an increasingly central role in assisting developers in their software creation activities. This is in essence a new form of software reuse in which collections of previously created software artifacts form the basis for generating new ones. Unlike in the past when developers were manually searching for pre-existing software components from libraries and code repositories such as Github, Node Package Manager (NPM) or the Python Package Index (PyPI), in the new model developers are requesting AI-driven assistants to generate suitable pieces of code for them.  These generated artifacts can range from small code snippets and module fragments to comprehensive application skeletons or in some cases fully functional applications or even complete end-to-end systems.  

This new generative approach to software reuse has resulted in a considerable mental model change for developers.  Basically, developers must be capable enough to judge the quality and suitability of the artificially generated code to the task at hand.  This can in many cases be very challenging, since AI-driven assistants will likely use numerous different codebases to derive the automatically generated designs.  Furthermore, the generated code may no longer reflect the mental model of any human developer; rather, it can be an amalgamation of various different development styles that have been conceived from several unrelated but seemingly similar solutions.  Generated code can in many cases be nearly perfect for the intended purpose.  However, occasionally AI will hallucinate results that will look deceivingly convincing to an untrained eye but that are actually completely erroneous and bogus. 

In many ways, generative reuse can be viewed as a new form of \emph{cargo cult programming}, or inclusion of code or program structures that originates from external sources without consideration or adequate understanding of relevance or side-effects \cite{feynman1998cargo,maki2019cargo,taivalsaari2019programming,CargoCult2004}. Just like in cargo cult programming, developers are (re)using code that they do not really necessarily understand at all.  In the classic cargo cult programming scheme, developers are blindly doing something simply because others have used a certain piece of code or certain development approach earlier -- basically putting their trust on artifacts that have already been known to work in other contexts.  In contrast, in AI-assisted generative reuse developers place their trust on code that is generated by an external "oracle" whose inner workings are usually completely unknown to the user.

In this paper, we discuss the implications of this new model of generative reuse, as well as define a tentative research agenda and call for action for tackling some of the central issues associated with this emerging practice of "AI native" software engineering.

\section{Evolution of Software Reuse} \label{sec:bg}

Ever since software reuse became a software engineering discipline in the early 1970s, there has been a call for systematic software reuse practices. By definition, software reuse is the use of existing software or software knowledge to construct new software. The promise of reuse is systems that are larger and more complex -- yet potentially more reliable and less expensive -- by reducing redundancy in code and by taking advantage of assets that have already been created earlier. The key idea in reuse is that parts of a computer program written at one time can be used in the construction of other programs later \cite{frakes2005software}. Reusability is traditionally considered as one of the key \emph{"ilities"} or major software quality factors -- along with usability, maintainability, scalability, availability, extensibility and portability \cite{frakes2005software}. 

General purpose, commercially available software component libraries have been proposed ever since the famous NATO 1968 conference in which the term \emph{software engineering} was also introduced \cite{nato1968}. As a research topic, software reuse became especially popular in the 1980s \cite{lanergan1984software,lenz1987software,biggerstaff1987reusability}, following the successful workshop on software reuse arranged by ITT Corporation in September 1983. In practice, commercial success of large-scale software reuse and component libraries did not begin until the proliferation of the World Wide Web in the 1990s, though. 

The forefathers of software engineering identified four types of reuse: (1) \emph{ad hoc} or informal, (2) \emph{opportunistic}, (3) \emph{systematic}, and (4) \emph{institutionalized} reuse \cite{nato1968,krueger1992software,hartmann2008hacking}. The former two approaches were considered inferior and avoidable \cite{mili1995reusing}.  The strong pressure (and ultimately failure) to introduce systematic software reuse practices in the software industry was also the main reason why many thought in the 1990s that software reuse was effectively dead, at least as an academic research topic \cite{schmidt1999software}. In earlier research, a distinction was also made between \emph{code reuse}, \emph{design reuse}, \emph{specification reuse}, and \emph{application system reuse} \cite{Cheng1994}.

In the early days of software development, the amount of reuse remained remarkably low. Back in those days, developers took pride in developing as much of the code themselves as they could, and reuse was often seen as the last resort.  If reuse took place, it was usually allowed only when there was an opportunity to reuse code from truly reliable and trustworthy sources such as close colleagues and other recognized technology experts.

The low level of reuse at that time was in striking contrast with the fact that for decades it was known that only a relatively small portion of software systems was truly unique for a particular system under construction. For instance, back in 1984 Jones reported that on average only 15 percent of code is truly unique, novel and specific to individual applications, whereas the rest was considered common, generic, and concerned with making the program to cooperate with the surrounding execution environment \cite{jones1984reusability}. Other early studies carried out at that time reported varying reuse rates between 10 and 60 percent \cite{lanergan1984software,lenz1987software,biggerstaff1987reusability}. Still, despite the potential for software reuse, actual reuse rates remained low, and the developers were obliged to write their own code \cite{taivalsaari2019programming}. 

Quite obviously, prior to the widespread use of the World Wide Web, software reuse was limited by \emph{component distribution and discoverability challenges}. In the bygone era of shrink-wrapped software, large-scale distribution of components was possible realistically only for larger institutions.  Furthermore, those components were rarely available for free or with license terms that were fit for large-scale use.

It was really the World Wide Web that made mass distribution and reuse of software components possible for ordinary developers in a "friction-free" fashion \cite{taivalsaari2019programming}. Over time, the availability of such libraries grew exponentially, as nowadays evidenced by popular software repositories such as GitHub, Node Package Manager and Python Package Index that literally offer millions of components -- and most of them with very liberal, permissive open source license terms. 
As an anecdote, the total number of NPM packages exceeded 3.3 million in July 2023 according to \emph{State of NPM 2023} report \cite{StateOfNPM}. %

As a result, in a relatively short period of time, the software industry went from very limited amount of reusable components to a total cornucopia in which the main limiting factors were \emph{searchability} and \emph{trust} rather than availability of components \emph{per se}.

The plentiful availability of reusable open source assets for just about any domain has changed the nature of system development profoundly. Contrary to textbook examples (e.g., \cite{kim1998software}), developers rarely perform reuse in a planned and managed fashion, as assumed in product line development \cite{clements2002software}. Instead, developers reuse software in an \emph{opportunistic}, mix-and-match fashion, basically trawling and scraping the Internet repositories for most suitable or most easily available components, then combining and configuring those already existing solutions, and decorating them with relatively minor application-specific modifications and additions. 

This approach was all about combining unrelated, often previously unknown hardware and software artifacts by joining them with "duct tape and glue code" \cite{hartmann2008hacking}. Depending on one's viewpoint, desired connotation and domain, such development is referred to as \emph{opportunistic design} \cite{hartmann2008hacking}, \emph{opportunistic reuse},  \emph{ad hoc reuse}, \emph{scavenging} \cite{krueger1992software}, 
\emph{trawling} \cite{mikkonen2010mashware}, or sometimes even \emph{frankensteining} \cite{hartmann2008hacking}.  

Such reuse is spontaneous by nature and driven by immediate needs, often arising in response to specific challenges or time constraints.  The resulting systems resembled \emph{icebergs}, with only the "tip of the iceberg" written by developers themselves, whereas the bulk of the system comes from other sources and remains invisible and poorly understood under the water \cite{taivalsaari2019programming}. Quite often, the developer has no idea of what code or how much code they are actually reusing, since the included components may dynamically pull in hundreds or even thousands of additional subcomponents. 

The iceberg and trawler analogies are accurate also in the sense that developers themselves often feel "lost at sea" when it comes to decisions as to what components or program structures to (re)use.  Online resources such as StackOverflow, Reddit, JavaRanch and many others have turned out to be invaluable for developers in guiding them through treacherous waters towards rational technology and component choices \cite{makitalo2020opportunistic}.

The resulting approach bears the imprint of \emph{cargo cult programming} or \emph{cargo cult development} \cite{feynman1998cargo,maki2019cargo,taivalsaari2019programming,CargoCult2004} -- the ritual inclusion of code or program structures for reasons that the programmers do not necessarily understand at all.  Such development practices have become surprisingly common given the massive amounts of code available in public repositories and time constraints in today's agile software development projects.  Increasingly often, component selection decisions are based simply on popularity ratings and the number of \texttt{"git pull"} requests in public code repositories.

\section{Towards AI Native Software Development}

In recent years, there has been a tremendous shift towards the use of AI in various domains -- not only in the software industry but in the society at large.  This boom towards AI-generated solutions has been very noticeable also in the software development area, where developers are increasingly relying on services such as ChatGPT, Gemini and CoPilot for generating software artifacts and for requesting advice on proposed programming patterns, development tools, libraries and methods \cite{siam2024programming}.  It is already rather obvious that the next logical step in the evolution of the software reuse continuum is to use AI -- and especially generative AI -- help developers create and reuse software. 

By \emph{AI native software development} (or \emph{AI-driven} or \emph{AI-assisted} software development), we refer to the use of AI tools and techniques to support, enhance, and automate various aspects of the software development lifecycle. This includes coding, debugging, testing, integration and deployment. AI models can provide real-time code suggestions, detect and fix errors, automate repetitive tasks, and even generate boilerplate code or entire modules based on natural language descriptions. By analyzing patterns from existing codebases and understanding programming context, these tools can help developers improve productivity, reduce errors, and accelerate development time. In essence, AI-assisted development acts as an intelligent collaborator, making the software creation process more efficient and accessible. In general, such techniques help less experienced developers act like true professionals, albeit there are also identified downsides \cite{ernst2022ai,waseem2024chatgpt}.

AI native software development relies on patterns learned from vast, diverse datasets of programs and documentation, including source code, documentation, tutorials, code snippets, and technical discussions (e.g., forums such as Stack Overflow) \cite{ernst2022ai}. Hence, it  essentially constitutes a new form of reuse, which has been widely adopted by developers, but which has received little attention from academic research despite its profound effect on software engineering practices.

It should be noted that generative reuse is not an entirely new concept.  There are existing survey papers from over twenty-five years ago (e.g. \cite{biggerstaff1998generative}), and roots of this approach can be traced back to 1980s. In his paper, Biggerstaff analyzed success factors and challenges in software reuse, evaluated a number of then known generative reuse approaches, as well as provided measurements and metrics related to their effectiveness. However, the concept of using AI for generating code and "weaving" entire solutions together from various existing sources is still relatively new.

\section{From Cargo Cult and Prompt Engineering to a Systematic Practice?} \label{sec:pcd}

At the time of this writing, it is still unclear what the long-term impact of AI native software reuse will be.  AI as a theme certainly has a cult-like following and "aura of omnipotence" around it at the moment -- people are putting a lot of faith on the ability of AI to magically accomplish almost anything. In those software development projects that we have been involved in the past 1 1/2 years, we have seen the use of AI tools increase dramatically, and it is very likely that the general approach is here to stay.

Speaking of expectations of omnipotence, at the moment a lot of people seem to have almost a religious belief in the "prompt engineering" capabilities of generative AI.  Having witnessed the power of such capabilities in areas such as automated image and video generation (including the use of Gaussian splatting techniques \cite{kerbl3Dgaussians} to generate vivid 3D scenes and even live videos out of conventional 2D photographs), a lot of people are assuming that in the area of software development it would ultimately be possible for AI to generate not only simple application skeletons but complete end-to-end software systems simply by providing some textual paragraphs describing the requested system.  In the ultimate scenario, traditional requirements definition phase would be skipped altogether based on the assumption that AI could "fill in the missing gaps" automatically.  In many ways, such prompt engineering would represent the ultimate form of cargo cult development. Relevant questions in this area include:

\begin{itemize}
\item Can AI realistically generate complete, fully functional end-to-end systems?
\item Can AI realistically help to comprehend generated code?
\item What are the realistic limits of "prompt engineering" in generating real life systems?
\item Is generative reuse an acceptable approach for generating code for production system use?
\item If the use of generative reuse becomes prevalent, can it be turned into a systematic method instead of the current \emph{"let's try and see what AI can come up with"} approach?
\item Is it possible to define systematic practices for generative AI-assisted reuse, or are the notions of a systematic method and generative AI fundamentally in conflict?
\item What is the equivalent of institutional reuse in the context of artificially generated code?
\item How do we train developers to evaluate the quality of artificially generated code?
\item How can developers evaluate quality of the generated artifacts if there are no natural subsystem boundaries or if the generated code does not follow any typical development patterns used by human developers?
\item How do we address, e.g., the copyright and security concerns associated with generative AI-assisted reuse?
\item How do we deal with maintenance, documentation, and other aspects related to reuse activities that are not directly associated with software generation and delivery?
\end{itemize}

Given that the innards of generative AI systems are effectively sealed "mystery boxes" to the users, the use of generative AI-driven reuse solutions poses various challenges.  Basically, the developers using AI-generated code must be skillful enough to distinguish and evaluate the quality of the generated artifacts. While this may still be possible when the generated amounts of code are small, the situation can become unmanageable even for experienced software developers when entire end-to-end systems are being generated. Merely blindly trusting whatever was automatically generated cannot be an acceptable approach if or when millions of lines of code are generated without human oversight or reviews.  Key concern here is that reuse based on assets that have not been proven to be fit for purpose or designed with reuse in mind can lead to serious quality issues. This is in sharp contrast with more classical forms of reuse that are assumed to improve software quality \cite{gao2003testing,trendowicz2003quality}.

Already back in 1997, Biggerstaff noted that an optimal strategy for generative reuse appears to favor systems in which there is a clear subsystem structure and in which those subsystems are generally organized in such a fashion that the size of an individual subsystem is in the KLOC range, i.e., size of individual subsystems and modules is only a few thousands of lines \cite{biggerstaff1998generative}.  AI-generated code might not follow such conventions; instead it could simply aim at a monolithic system that immediately satisfies the given requirements. This in turn can introduce a mismatch with approaches that are conventionally used in testing, debugging, and deployment.

Finally, it should be noted that with AI-assisted development,
developers are ultimately reusing fragments of the training datasets. Therefore, developers should keep their eyes open for not only direct but "tangential" copyright and license term violations; while generated code might not be 100\% identical with known sources, it might still be close enough to trigger copyright and license issues. Besides quality and safety concerns, copyright issues are generally an interesting concern, since the ownership rights of AI-generated code are not always entirely clear.  Nowadays, there are tools such as \emph{Black Duck SCA} (see \url{https://www.blackduck.com/}) to detect such indirect copyright and license violations.

\section{Call to Action} \label{sec:cta}

Over the past decades, the software industry has progressed from classic small-scale software reuse and aspirations of systematic software reuse practices to the recent era of opportunistic reuse enabled by the massive proliferation of available open source components
\cite{hartmann2008hacking}.  This trend is currently being disrupted by generative artificial intelligence, which seemingly increases developer performance by a considerable margin \cite{li2024sheetcopilot}. 

Whether we like it or not, AI native software development has arrived and is likely to have a significant impact on how software will be created in the future -- if not to replace the earlier software development and reuse practices altogether.  As we have argued in this paper, AI native software reuse in its current form can be viewed as an extension or continuum of the opportunistic reuse approach that became popular in the past decade.  In many ways, AI native software development represents the ultimate form of the "tip of the iceberg" reuse model; in the extreme scenario developers do not write any of the code themselves; practically speaking all of the code in such a system would remain "under the water" and poorly understood by the developers. 

So far, the emergent use of generative AI approaches in software development has received relatively little attention among researchers focusing on software reuse.  Since using generative AI to assist software design and development is becoming prevalent among software developers, the software engineering community must start seriously studying and considering the consequences and challenges associated with the approach, resulting in what we have referred to as AI native software engineering in this paper.

Effectively, the questions that we have listed in the previous section form the beginning of a research agenda that could be used for turning generative reuse from a mere variant of cargo cult development into a more systematic practice. While it is still difficult to predict the long term impact of the disruption caused by AI native software development, we hope that this paper -- for its part -- encourages researchers and the broader software engineering community to actively work towards tackling the issues associated with this emerging (and admittedly exciting) approach.

\bibliographystyle{plain}
\bibliography{sample-base}

\end{document}